\documentclass[12pt]{article}
\usepackage[scaled]{helvet}
\usepackage[margin=1in]{geometry}
\RequirePackage[OT1]{fontenc}
\RequirePackage{amsthm,amsmath}

\usepackage{enumerate}
\usepackage{natbib}
\usepackage{url}
\usepackage{amsmath}
\usepackage{graphicx,psfrag,epsf}
\usepackage{booktabs}
\usepackage{amssymb}
\usepackage[margin=1in]{geometry}
\usepackage{setspace}
\usepackage{subcaption}
\usepackage{amsfonts}
\usepackage{titlesec}

\titleformat*{\section}{\normalsize\bfseries}
\titleformat*{\subsection}{\normalsize\it}
\titleformat*{\subsubsection}{\normalsize\it}
\usepackage[affil-it]{authblk}

\graphicspath{img/}

\title{Treed distributed lag nonlinear models}

\author[1]{Daniel Mork\footnote{Daniel.Mork@colostate.edu}}
\author[1]{Ander Wilson\footnote{Ander.Wilson@colostate.edu}}

\affil[1]{Department of Statistics, Colorado State University, Fort Collins, CO, USA}
\date{\vspace{-2cm}}

\begin{document}
\maketitle
\setstretch{1.8}

\noindent\textbf{Summary}\\
In studies of maternal exposure to air pollution a children’s health outcome is regressed on exposures observed during pregnancy. The distributed lag nonlinear model (DLNM) is a statistical method commonly implemented to estimate an exposure-time-response function when it is postulated the exposure effect is nonlinear. Previous implementations of the DLNM estimate an exposure-time-response surface parameterized with a bivariate basis expansion. However, basis functions such as  splines assume smoothness across the entire exposure-time-response surface, which may be unrealistic in settings where the exposure is associated with the outcome only in a specific time window. We propose a framework for estimating the DLNM based on Bayesian additive regression trees. Our method operates using a set of regression trees that each assume piecewise constant relationships across the exposure-time space.  In a simulation, we show that our model outperforms spline-based models when the exposure-time surface is not smooth, while both methods perform similarly in settings where the true surface is smooth. Importantly, the proposed approach is lower variance and more precisely identifies critical windows during which exposure is associated with a future health outcome. We apply our method to estimate the association between maternal exposure to PM$_{2.5}$ and birth weight in a Colorado USA birth cohort.


\section{Introduction}
\label{sec1}
In many applications there is interest in regressing an outcome on exposures observed over a previous time window. This frequently arises in environmental epidemiology applications where either a health outcome on one day is regressed on exposures (e.g. temperature or air pollution) observed on that day and several proceeding days or when a birth or children’s health outcome is regressed on exposures observed daily or weekly throughout pregnancy \citep{Stieb2012}. 

In the context of maternal exposure to air pollution, which we consider in this paper, there are generally two inferential goals. The first is to estimate the critical windows of susceptibility--periods in time during which an exposure can alter a future phenotype.  The second goal is to estimate the exposure-time-response function. Recent studies have identified critical windows and associations between maternal exposure to air pollution and several outcomes including preterm birth \citep{Chang2012,Chang2015a} adiposity \citep{Chiu2017}, asthma and wheeze \citep{Bose2018a,Lee2017}, neurodevelopment \citep{Chiu2016a}, among other outcomes \citep{Stieb2012,Sram2005}. This includes studies that have found that the linear \citep{Chiu2017, Chang2015a} and nonlinear \citep{Wu2018a} association vary across weeks of gestation.

A popular approach to estimate the association between maternal exposure to air pollution during pregnancy and a birth outcome is a distributed lag model (DLM) \citep{Schwartz2000a,Zanobetti2000}. In a DLM, the outcome is regressed on the exposures at each of the time points, e.g. mean exposure during each week of pregnancy. Most commonly, the model is constrained so that the exposure effect varies smoothly over time. Constraining the model adds stability to the estimator in the presence of typically high temporal correlation in the exposure. Methods of regularization include penalized spline regression \citep{Zanobetti2000}, Gaussian processes \citep{Warren2012}, principal components or splines \citep{Wilson2017a}.  \cite{Wilson2017} showed that a constrained DLM outperforms more naive methods such as using average exposure over each of the trimesters because DLMs adjust for exposures at other time points throughout pregnancy and provide a data driven approach to identify critical windows even when they do not align with clinically defined trimesters.

To extend the DLM to estimate nonlinear associations in the exposure-response function at any given time, a class of distributed lag nonlinear models (DLNMs) has been proposed \citep{Gasparrini2010, Gasparrini2017}. DLNM methods typically operate by cross-basis smoothing with splines or penalized spline regression. This results in a unique nonlinear exposure-response function at each time point that varies smoothly over the lagged exposures.

A consequence of imposing smoothness over time in a DLM or DLNM is that estimates may generalize the critical window(s) to a wider set of times than is appropriate. Critical windows are hypothesized to be defined by biological events in the fetal developmental process that may be altered by environmental exposures. Methods that can adapt to the discrete time spans of these events are needed to better estimate critical windows. Motivated by this, \cite{Warren2019} proposed a hierarchical Bayesian framework to improve critical window characterization for DLMs using a variable selection approach that selected weeks in or out of the critical window. However, there are no DLNM methods that relax the smoothness constraint for a nonlinear exposure-response function.

In this paper, we propose a method for DLNM that relaxes the smoothness assumption and can more precisely identify critical windows. The proposed approach, which we call treed distributed lag nonlinear models (TDLNM), is based on the Bayesian additive regression trees (BART) framework developed by \cite{Chipman2012}. Applied to estimating a distributed lag function, TDLNM treats the time series of exposures as a single multivariate predictor and uses a tree structure to partition the exposure concentration and time dimensions to construct a flexible exposure-time-response surface.

We propose two forms of TDLNM. The first form uses a dichotomous tree structure to form a piecewise constant exposure-time-response function. By using an ensemble of trees, the model can approximate both smooth and non-smooth functions. The second form imposes smoothness only in the exposure-concentration dimension but not over time. This forces smoothness in the exposure-response, while maintaining precision in critical window identification. We also discuss how the smooth version can be used to incorporate exposure uncertainty into the model.

Following development of TDLNM, we perform a simulation study that compares our proposed method to spline-based methods across a variety of settings. These simulations demonstrate that our method excels in the estimation of the exposure-time-response function for non-smooth settings, but also adapts well to estimating scenarios with a smooth exposure-time-response. Importantly, we find that TDLNM more precisely identifies critical windows and has an extremely low rate of critical window misspecification. In addition, simulations show that TDLNM has narrower confidence intervals, especially near the boundaries, while maintaining nominal coverage. Finally, we apply TDLNM to estimating the association between the fine particulate matter (PM$_{2.5}$) experienced by a mother during pregnancy and the resulting birth weight of the child. Software to implement this method is available in the {\tt R} package {\tt dlmtree}.

\section{Data}

We analyze birth records from Colorado, USA, vital statistics data. The data includes live, singleton, full term ($\ge 37$ weeks gestation) births from Colorado with estimated conception dates between 2007 and 2015, inclusive, with no known birth defects. We limited the analysis data to the northern front range counties (those immediately east of the Rocky Mountains roughy extended from Colorado Springs to the Wyoming border). This area contains the majority of the Colorado population. We further limited the analysis to census tracts with elevation less than 6000 feet above sea level. This both reduces the potential confounding by altitude and the impact of mountainous terrain on exposure data.

The primary outcome of interest in this paper is birth weight for gestational age $z$-score (BWGAZ). We obtained BWGAZ using the Fenton birth charts \citep{Fenton2013a}. BWGAZ measures birth weight as the number of standard deviations above or below the expected birth weight of a child with the same fetal age and sex. The data contain individual level covariate information including mother's age, weight, height, income, education, marital status, prenatal care habits and whether they smoked before or during pregnancy, as well as race and Hispanic designations. We limit the analysis to observations with complete covariate information, resulting in 300,463 births.

We use PM$_{2.5}$ exposure data from the US Environmental Protection Agency fused air quality surface using downscaling data. This data is publicly available at \url{www.epa.gov/hesc/rsig-related-downloadable-data-files}. The statistical methodology for construction of the data files has been described in \cite{Berrocal2009AModels}. We linked the exposure data to the birth records based on the census tract of maternal residence at birth. We then constructed weekly average exposures for each week of gestation. A map detailing the number of births in each county is shown in Supplemental Figure 1.

This study was approved by the Institutional Review Board of Colorado State University.

\section{Methods}

\subsection{DLNM Framework}
Before introducing our proposed method we briefly recap the DLNM framework and standard methodology. Let $y_i$ be the continuous outcome for person $i$ from a sample $i=1,\dots,n$. Let $\mathbf{x}_i=[x_{i1},\ldots,x_{iT}]^T$ denote a vector of exposures observed at equally spaced times $t=1,\ldots,T$. In our case, $y_i$ indicates BWGAZ while $x_{it}$ represents the $i^{th}$ mother's exposure to PM$_{2.5}$ in week $t$ of pregnancy. We control for a vector of covariates, denoted $\mathbf{z}_i$. The Gaussian DLNM model is 
\begin{equation}
\label{eq:main}
    y_i\sim\mathcal{N}\left[f(\mathbf{x}_i)+\mathbf{z}_i^T\boldsymbol{\gamma}, \sigma^2\right],
\end{equation}
where $f(\mathbf{x}_i)$ is the distributed lag function and $\boldsymbol{\gamma}$  is a vector of regression coefficients.  

The distributed lag function $f(\mathbf{x}_i)$ can take several linear as well as nonlinear forms. The DLNM allows a unique nonlinear association between exposures at each time point and the outcome. In general, the distributed lag function is defined as
\begin{equation}\label{eq:dlnm}
    f(\mathbf{x}_i)=f(x_{i1},\ldots,x_{iT})=\sum_{t=1}^T w(x_{it},t)
\end{equation}
where $w(x,t)$ is the exposure-response function relating exposure at week $t$ of gestation to the outcome. Existing frameworks for the DLNM \citep{Gasparrini2010,Gasparrini2017} utilize a cross-basis where $w$ is represented as a bivariate basis expansion in the exposure concentration and time dimensions. Penalized spline implementations allow for a range of assumptions to be made regarding the structure of the exposure-time-response. For example, varying ridge penalties target shrinkage at specific times, while varying difference penalties control the smoothness along the curve. Basis expansion methods, such as splines, regularize the model to improve stability of the estimated effect in the presence of multicollinearity in the predictor. However, these methods also impose the assumption of smoothness in the DLNM.

\subsection{Treed DLNM Approach}\label{sec:treed-dlnm}
\begin{figure}[!p]
\begin{subfigure}{.5\textwidth}
  \centering
  \includegraphics[width=.8\linewidth]{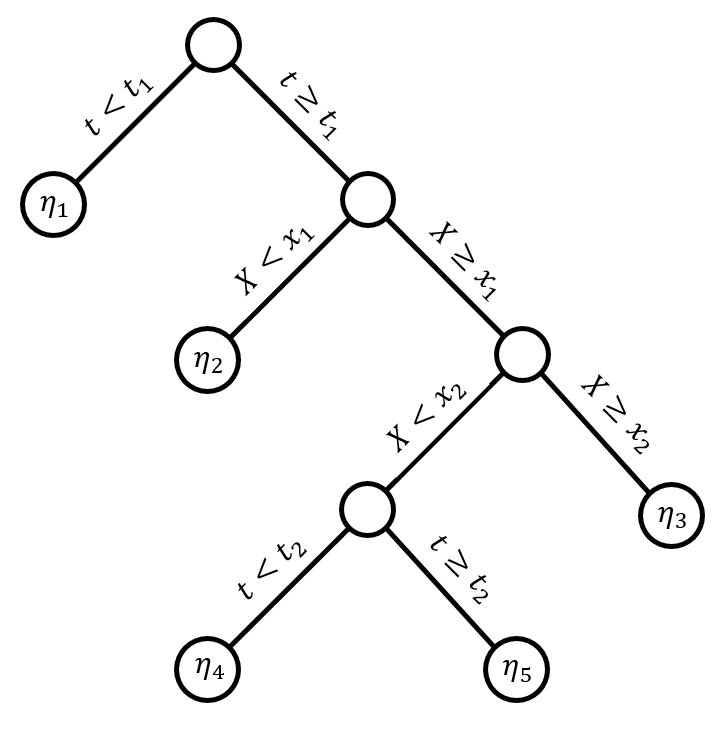}  
  \caption{}
  \label{fig:dlnm-tree-example-a}
\end{subfigure}
\begin{subfigure}{.5\textwidth}
  \centering
  \includegraphics[width=.8\linewidth]{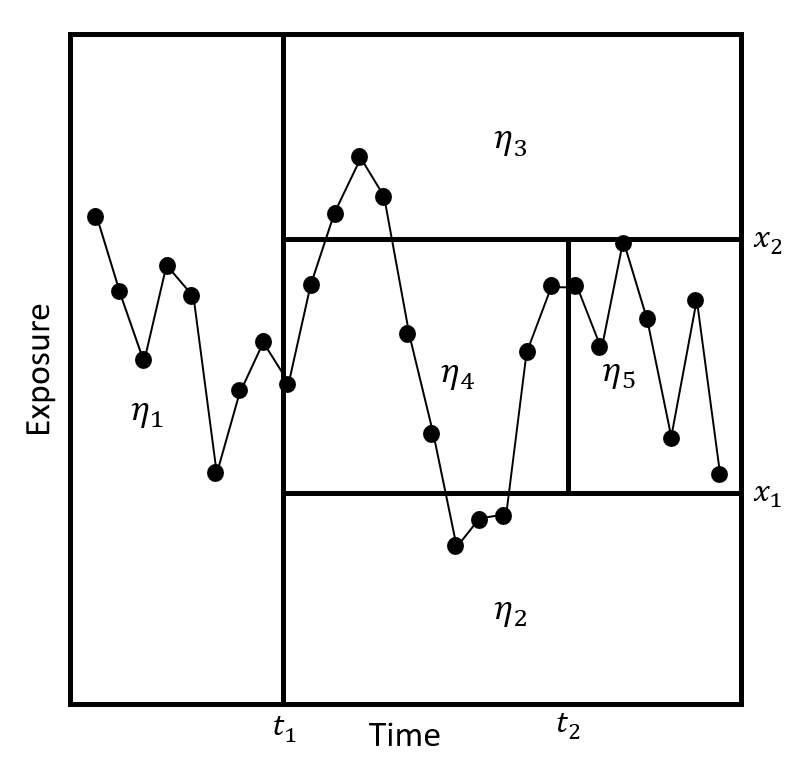}  
  \caption{}
  \label{fig:dlnm-tree-example-b}
\end{subfigure}
\caption{Example of tree, $\mathcal{T}$, with terminal nodes $\eta_{b}$, $b\in\{1,2,3,4,5\}$. Panel (a) diagrams the tree with dichotomous splits on time or exposure concentration while panel (b) represents the resulting partition of the exposure-time space for a single observation.}
\label{fig:dlnm-tree-example}
\end{figure}

We introduce a sum-of-trees model based on the BART framework \citep{Chipman2012} to estimate the exposure-time-response function, $f(\mathbf{x}_i)$. The general approach is to build dichotomous trees that partition the time-varying exposure $\mathbf{x}_i$ in both the exposure concentration and time dimensions. Figure~\ref{fig:dlnm-tree-example} illustrates the approach for a single tree. Figure~\ref{fig:dlnm-tree-example-a} is a diagram of a tree showing binary rules defined on the exposure and time values. These rules divide the exposure-time space into five terminal nodes, denoted $\eta_1,\ldots,\eta_5$. Figure~\ref{fig:dlnm-tree-example-b} shows the exposure-time space partitioned into five regions with each region corresponding to a single terminal node. A tree and corresponding parameters define a piecewise constant exposure-response function,
\begin{equation}
\label{eq:wa_constant}
    w(x_{it},t)=\mu_{b} \quad \text{if } x_{it}  \in  \eta_{b}.
\end{equation}
The distributed lag function for tree $\mathcal{T}$ takes a form similar to that in \eqref{eq:dlnm} and is defined as
\begin{equation}
\label{eq:partial-dlnm}
    g(\mathbf{x}_i,\mathcal{T})= \sum_{t=1}^T w(x_{it},t).
\end{equation}

In our TDLNM framework, we consider an ensemble of $A$ regression trees. For tree $\mathcal{T}_a$, $a\in\{1,\ldots,A\}$, denote the $B_a$ terminal nodes as $\eta_{a1},\ldots,\eta_{aB_a}$. Each terminal node, $\eta_{ab}$ has a corresponding set of limits in time and exposure concentration given by the rules defined as splits of the tree and a corresponding parameter $\mu_{ab}$. Collectively, the terminal nodes of each tree define a partition of the exposure-time-space and allows for flexible estimation of the exposure-time surface. As in \eqref{eq:wa_constant}, we define the effect of each exposure-time combination in tree $\mathcal{T}_a$ to be $w_a(x_{it},t)=\mu_{ab}$ if $x_{it}\in\eta_{ab}$. Each regression tree in the ensemble provides a partial estimate of the distributed lag nonlinear function $f$. Formally, the exposure-time-response function for TDLNM is
\begin{equation}
\label{eq:partial-dlf}
    f(\mathbf{x}_i)=\sum_{a=1}^A g(\mathbf{x}_i,\mathcal{T}_a),
\end{equation}
where $g(\mathbf{x}_i,\mathcal{T}_a)$ represents the partial estimate contributed by tree $a$ given in \eqref{eq:partial-dlnm}.

TDLNM foregoes the basis-imposed smoothness assumption. However, when different time and exposure breaks are staggered across trees the ensemble of trees can approximate smooth functions. Model regularization is a result of the tree prior, which prefers trees having only a few splits. Smaller trees ensure that the model is stable in the presence of temporal correlation because each terminal node averages across multiple time points.

\subsection{Smoothing in exposure concentration} \label{sec:smoothing_exposure_conc}
Most epidemiological studies assume that the exposure-response relationship is smooth in exposure concentration. The TDLNM methods presented above assume a piecewise linear structure that can approximate a smooth function, but it is never truly smooth. In this subsection we propose a TDLNM model that is truly smooth in exposure (TDLNMse). Importantly, TDLNMse does not force smoothness in time to allow for accurate critical window estimation.

To allow for smoothing in the exposure-response, we introduce a weight function on the terminal-node specific effects. A similar idea was introduced by \cite{Linero2018}, which assigned a node-specific probability to each observation using a gating function at each dichotomous split on a covariate. TDLNMse differs in that we desire smoothing only in the exposure-concentration dimension. To accomplish this, we define smoothing parameter $\sigma_x$ and modify \eqref{eq:wa_constant} to be
\begin{equation}
    \label{eq:wa_smooth}
    w_a\left(x_{it},t\right)=
    \sum_{b=1}^{B_a} \mu_{ab}\cdot \psi(x_{it}; \eta_{ab},\sigma_x).
\end{equation}
The weight function $\psi(x_{it}; \eta_{ab},\sigma_x)$ allows each observation $x_{it}$ to be distributed across all terminal nodes that contain time point $t$. For the weight function we use a normal kernel with bandwidth $\sigma_x$. Hence, the weight for $x_{it}$ assigned to node $\eta_{ab}$ is
\begin{equation}\label{eq:weight_x_in_eta}
    \psi(x_{it}; \eta_{ab},\sigma_x)=
    \left\{
        \Phi\left(\frac{\lceil x_{ab}\rceil-x_{it}}{\sigma_x}\right)-
        \Phi\left(\frac{\lfloor x_{ab}\rfloor-x_{it}}{\sigma_x}\right)
    \right\}
    \cdot \mathbb{I}(t\in\eta_{ab}),
\end{equation}
where $\lceil x_{ab}\rceil$ and $\lfloor x_{ab}\rfloor$ refers to the maximum and minimum exposure concentration limits of node $\eta_{ab}$, respectively, and $\Phi$ is the standard normal cumulative distribution function. The inclusion of the indicator function allows TDLNMse to retain a piecewise constant effect in time at each exposure concentration value. The kernel smoother requires fewer terminal nodes to estimate a smooth effect in exposure-concentration as observations near the boundary of two terminal nodes will have an estimated effect that is in between estimates of observations located centrally in the nodes.

For TDLNMse, we propose fixing the bandwidth $\sigma_x$ a priori. Alternatively, we could assign a prior to $\sigma_x$ and estimate the bandwidth.

\subsection{Incorporating exposure uncertainty with TDLNMse} \label{sec:incorporating_measurement_error}
A situation that has not been addressed in the DLNM literature is uncertainty in the exposure. Many exposure models including climate models and  spatially kriged exposure models provide measurement of uncertainty. Most commonly these occur in one of two forms---standard errors for the exposure data or multiple realizations from a model such as draws from a posterior predictive distribution or an ensemble method. This uncertainty is not accommodated for in the health effect estimates from standard DLNMs. 

Exposure uncertainty can be incorporated into TDLNM by using a weight function to spread the exposure across multiple terminal nodes according to the probability that the exposures is in each of those nodes. The result is similar to TDLNMse, using a weight function corresponding to the uncertainty in each observation. In the case of reported standard errors for the exposure data, we use \eqref{eq:weight_x_in_eta} with observation specific smoothing parameter $\sigma_{xi}$ that is equal to the standard error for each observation. If instead we have multiple draws of exposures from an ensemble or Bayesian model we replace $\boldsymbol\Phi$ in \eqref{eq:weight_x_in_eta} with the empirical cumulative distribution function.

\subsection{Interpretation of TDLNM and relation to spline-based DLNMs}

To gain some insight into the exposure-response function characterized by TDLNM we consider the DLNM relation at a single time point. The distributed lag function in TDLNM is given by combining \eqref{eq:partial-dlnm} and \eqref{eq:partial-dlf}, i.e. the sum over trees and the sum of each tree over time. Reversing the order of the summation we get 
\begin{equation}
\label{eq:TDLNM-dl-est}
    f(\mathbf{x}_i)=\sum_{t=1}^T\sum_{a=1}^A  w_a(x_{it},t).
\end{equation}
At time $t$, the exposure-response function, $\sum_{a=1}^A w_a(x_{it},t)$, is  equivalent to the BART model with univariate predictor $x_{it}$. In the case of TDLNM this implies a piecewise-constant exposure-response function across the exposure concentration levels at time $t$. For TDLNMse, the weight function $\psi$ acts as a linear smoother over exposure concentration for the exposure-response function at each time.

\section{Prior Specification and Computation}

Our prior specification is based on that of \cite{Chipman2012}; however, some modifications and a different MCMC algorithm are needed to accommodate or improve performance with the multivariate predictor and parametric control for covariates. In this section we specify key differences in the priors and computation approach from that of \cite{Chipman2012}, including a horseshoe-like shrinkage prior on tree-specific effects and an altered prior for tree splits on a multivariate predictor. Full details on the priors and computation are in the Supplemental Materials, Section B.

\subsection{Prior Specification}
We apply a tree-specific, horseshoe-like prior to the effects at the terminal nodes $\mu_{ab}$ \citep{Carvalho2010}. The prior for terminal node $b$  on tree $a$ is
\begin{equation}\label{eq:dlnm-prior}
\mu_{ab}|\sigma^2,\omega^2,\tau_a^2 
\sim
\mathcal{N} \left(0, \sigma^2\omega^2\tau_a^2\right).
\end{equation}
Here $\tau_a\sim\mathcal{C}^+(0,1)$ and $\omega\sim\mathcal{C}^+(0,1)$ defines the horseshoe prior on trees. We specify prior $\sigma\sim\mathcal{C}^+(0,1)$ and $\boldsymbol{\gamma}\sim \mathcal{MVN}(\boldsymbol{0},\sigma^2 c\mathbf{I})$, where $c$ is fixed at a large value.

For the half-Cauchy priors on all variance parameters we adopt the hierarchical framework of \cite{Makalic2015}, where $r^2|s\sim \mathcal{IG}(1/2,1/s)$ and $s\sim\mathcal{IG}(1/2,1)$ gives that marginally $r\sim\mathcal{C}^+(0,1)$. This allows for Gibbs sampling of all variance components.

The tree-specific shrinkage prior on $\mu_{ab}$ results in better mixing throughout the MCMC sampler. This occurs by allowing shrunken trees with a small variance component and smaller effects $\mu_{a1},\ldots,\mu_{aB_a}$ to more easily explore splitting locations in the exposure-time space. After reconfiguration, these trees have the ability to contribute larger partial estimates.

Our stochastic tree generating process largely follows \cite{Chipman1998}. The probability a tree splits at node $\eta$ with depth $d_\eta$ equals $p_\text{split}(\eta)=\alpha(1+d_\eta)^{-\beta}$, where hyperpriors $\alpha\in (0,1)$ and $\beta\geq 0$. In our data setting the number of potential splits in the time direction is $T-1$ while the number of potential split points in the exposure direction is equal to the number of unique exposure values minus one, which is substantially larger than $T-1$. To address this imbalance we limit the potential exposure split points a priori and propose an alternative prior on potential split points. By limiting the exposure split points we also avoid situations where a split in one dimension limits future splits in another dimension due to empty nodes. For example, if TDLNM has a tree that splits on an extreme value in the exposure-concentration dimension, it may be unable to further split on the time dimension due to lack of data in one partition of the exposure-time space. We restrict the potential split locations in the exposure dimension to a predefined set of quantiles or values. Specification of potential splitting values also improves computational efficiency by allowing for precalculation of counts or weights for the limited number of potential splits.

We assign prior probabilities uniformly across potential time splits and uniformly across potential exposure splits such that there is a $0.5$ probability of selecting either a time or exposure split as the first splitting rule in a tree. Hence, for $s_x$ and $s_t$ total potential exposure and time splitting values, respectively, the probability of the first split in a tree being on exposure equals $1/(2s_x)$ or $1/(2s_t)$ if the split is on time. For a splitting decision further down the tree, the splitting rule probability is proportional to the probabilities of the potential remaining splits in a selected node. Following a split in time, there are fewer potential remaining splits in time, increasing the probability that the next split will take place in the exposure dimension.

\subsection{MCMC Sampler}\label{sec:mcmc}
We estimate TDLNM using MCMC. The MCMC approaches used for BART do not apply to the current model for two reasons. First, the algorithm of \cite{Chipman2012} relied on the fact that any specific vector of predictors $\mathbf{x}_i$ is contained in a single terminal node on each tree, whereas TDLNM divides the exposures related to each observation across the terminal nodes. Second, we modify the algorithm to allow for parametric control of the confounding variables $\mathbf{z}$. Due to these differences we propose an alternative MCMC approach for the TDLNM model. In particular, we integrate out $\boldsymbol{\gamma}$ using standard analytical techniques. Then, we apply Bayesian backfitting \citep{Hastie2000} to simultaneously estimate the effects of the partial exposure-time-response based on the partition defined by each tree, $\mathcal{T}_a$. Full details of the MCMC sampler can be found in Supplemental Materials Section B.

\subsection{Hyperprior selection and tuning}
\label{sec:prior_setup}
Tree splitting hyperpriors were set to the defaults used in \cite{Chipman2012} with $\alpha=0.95$ and $\beta=2$; different settings did not improve results. Trees in TDLNM explore only two dimensions, which requires fewer trees to adequately explore the predictor space.  In preliminary work, we found that 10 to 20 trees was sufficient. Results did not change using more than 20 trees. We used $A=20$ trees for our simulation and data analysis. We assigned the stochastic tree process to grow or prune with probability 0.3 each and change with probability 0.4. The fixed smoothing parameter in TDLNMse, $\sigma_x$, is data dependent: too large and the estimated effect will appear linear, too small and the model reverts to TDLNM (no smooth effect). For our simulation and data analysis, we set $\sigma_x$ to half the standard deviation of the exposure data. We found this setting to balance a smooth effect while also clarifying nonlinearity in the exposure-concentration effects.

\subsection{Estimating the exposure-time-response function}\label{sec:estimating-dlnm}
The distributed lag nonlinear function $f$ includes the model intercept. To ease interpretation we remove the intercept by centering $f$ at a reference exposure value, $x_0$, at each time. As $f$ is estimated as a sum of tree, we center each tree at the reference value and we use the centered trees for posterior inference.

\section{Simulation}
We conduct a simulation study to compare the empirical performance of TDLNM and TDLNMse to established DLNM methods that use penalized and unpenalized splines. Key to the simulation is that we compare performance on simulation scenarios representing both smooth and non-smooth exposure-time-response functions.

We simulate data according to \eqref{eq:main} and \eqref{eq:dlnm}, using a sample size equivalent to our exposure data ($n=300,463$). To accurately represent the autocorrelation found in air pollution exposure data, we use PM$_{2.5}$ exposures from our data analysis. We simulate the DLNM surface using 37 consecutive weeks from each observation to simulate a full-term pregnancy.  We consider four simulated exposure-time-response functions. Each corresponds to a different true model (TDLNM, TDLNMse, smooth DLNM with splines, and linear DLM). The four DLNM scenarios are: A) piecewise constant effect in exposure across weeks $11-15$; B) linear effect in exposure across weeks $11-15$; C) smooth, nonlinear effect (logistic shape) in exposure across weeks $11-15$; D) smooth, nonlinear effect (logistic shape) in exposure with a smooth effect in time peaking at week $13$ and extending approximately five weeks in either direction. We generate the outcomes using log-transformed exposure data. All scenarios are centered at log-exposure value 1. Several cross-sections of the exposure-time surfaces are shown in Figures \ref{fig:sim-slice-exp} and \ref{fig:sim-slice-time}. Algebraic details of the DLNM surface for each scenario and a graphic representation can be found in the Supplemental Materials Section C.1.

We generate a set of covariates (five standard normal, five binomial with probability 0.5) and corresponding coefficients from standard normal. We include a seasonal trend by using ozone data. Specifically, we add a random ozone effect for every $5^\text{th}$ week (5, 10, \ldots, 35), where ozone measurement at each time is centered mean zero, scaled to have standard deviation one and multiplied against a draw from $\mathcal{N}(0,\sigma^2=0.04)$. This allows for a different seasonal trend for each simulated dataset that is correlated with both the exposure, PM$_{2.5}$, and the outcome. We set the error variance $\sigma^2$ such that $\text{Var}[f(\mathbf{x}_i)]/\sigma^2=1/1000$ to represent a realistic signal to noise ratio and run 500 simulation replicates in each scenario. The simulation design can be reproduced with the {\tt R} package {\tt dlmtree}.

\subsection{Simulation estimators and comparisons}
TDLNM and TDLNMse used the prior settings described in section \ref{sec:prior_setup}. Thirty evenly spaced values ranging between the 0.01 percentile and the 99.9 percentile of all log-exposure value were designated as potential splits in the exposure-dimension. After a burn-in period of 5,000 iterations, we ran each model for 15,000 iterations, thinning to every tenth draw.

We compare TDLNM and TDLNMse to several spline-based penalized and unpenalized DLNM models. The models are described as follows with more detail given in \cite{Gasparrini2017}.
\begin{itemize}
    \item GAM: base model defined by penalized cubic $B$-spline smoothers of rank 10 in both exposure and time dimensions, with second-order penalties, estimated with REML;
    \item DLM: using GAM with a linear assumption in exposure concentration;
    \item GLM-AIC: optimal number of unpenalized, quadratic $B$-splines in both exposure and time dimensions (df 1 to 10) selected by minimizing AIC;
    \item GAMcr: defined by replacing the cubic $B$-spline basis in GAM with cubic regression splines and penalties on the second derivatives;
    \item GAM-exp: GAM, replacing the second-order penalties with a varying ridge penalty.
\end{itemize}
To assess model performance, we center the DLNM for each model at log-exposure value $1$ and evaluate the estimated DLNM over a grid of points. 

In each model we include all 10 simulated covariates as well as indicators for year and month to control for the additional seasonal trend. We log-transform the exposure concentration values to reduce skew in the exposure data and allow for equally spaced knots in the spline basis models. The decision to log-transform the response has no impact on TDLNM as the model will produce identical results with or without a log-transform; it does impact the smoothing with TDLNMse.

\subsection{Simulation Results}
Summary measures of model performance are shown in Table \ref{tab:sim-results}. 
Here, we compare each model by the root mean square error (RMSE) of the entire exposure-time surface and broken down to the RMSE within and outside the simulated critical windows. We also show the empirical coverage of 95\% confidence intervals along with average confidence interval width. In addition, the models are compared on the probability of identifying a non-zero effect across grid points inside the simulated critical window (TP), the probability of incorrectly placing a non-zero effect across grid points outside the simulated critical window (FP), and the precision of correct identification of a non-zero effect: TP/(TP+FP). We designate a non-zero effect in the true exposure-time surface as any effect outside of the interval from $-0.005$ to $0.005$ to account for scenario B and C, which have a non-zero effect everywhere between weeks 11-15 and scenario D, which has a non-zero effect everywhere. Figures 2 and 3 show cross-sections of the exposure-time-response surface using estimates from models TDLNM, TDLNMse, and GAMcr. A non-zero estimate in the plots indicates a change in the response for any observation with that particular time and exposure-concentration value.

\begin{figure}[!p]
\centering
\begin{subfigure}{.48\textwidth}
  \centering
  \includegraphics[width=\linewidth]{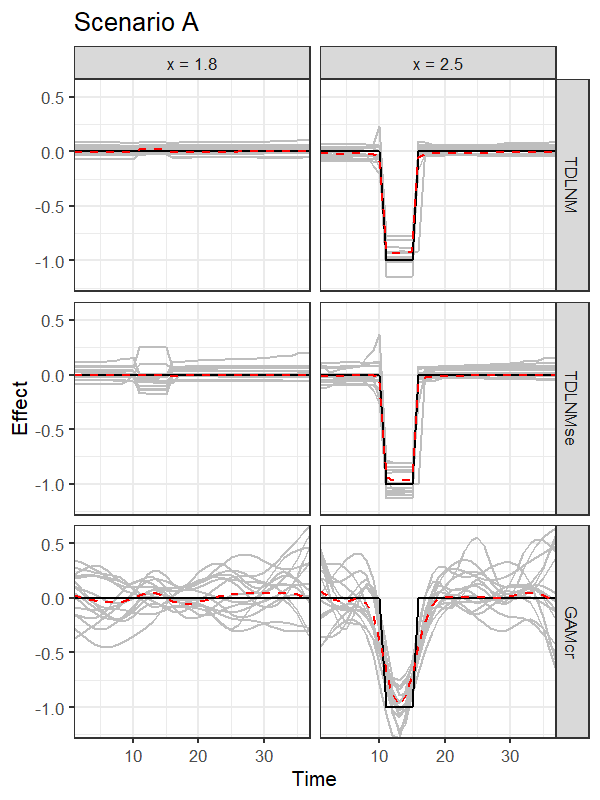}  
\end{subfigure}
\begin{subfigure}{.48\textwidth}
  \centering
  \includegraphics[width=\linewidth]{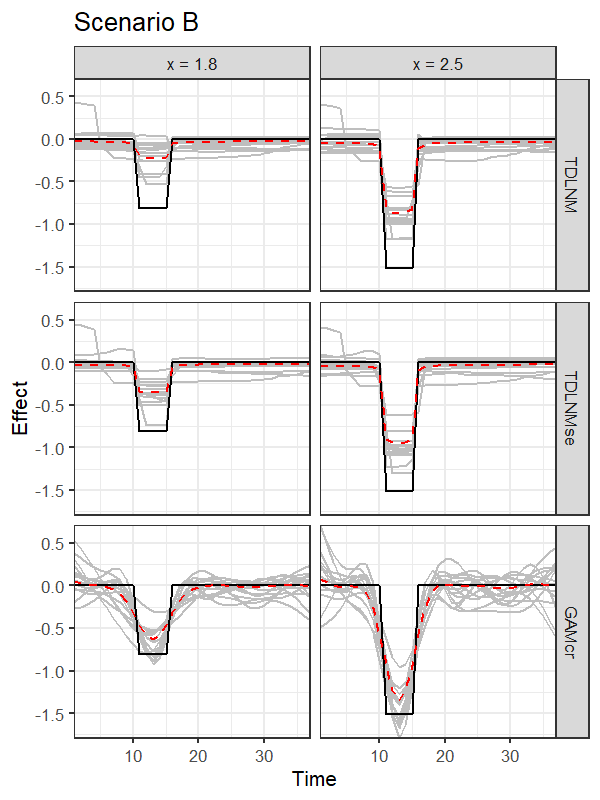}  
\end{subfigure}
\begin{subfigure}{.48\textwidth}
  \centering
  \includegraphics[width=\linewidth]{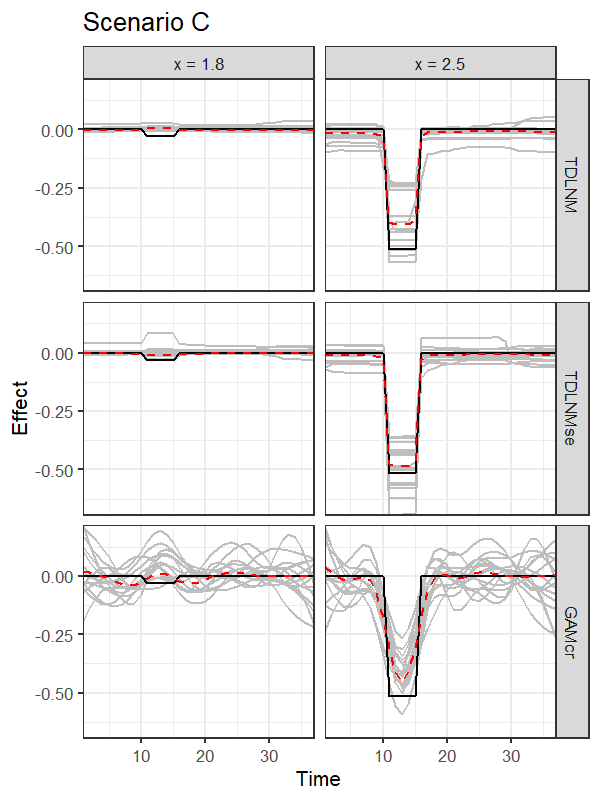}  
\end{subfigure}
\begin{subfigure}{.48\textwidth}
  \centering
  \includegraphics[width=\linewidth]{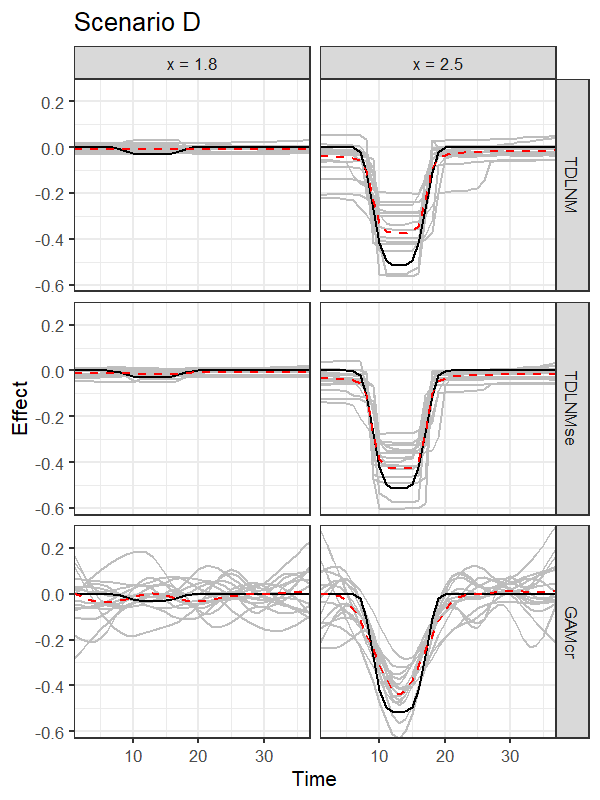}  
\end{subfigure}
\caption{Simulation results of TDLNM, TDLNMse, and GAMcr comparing the exposure-time-response (y-axis) cross-section across all times (x-axis) fixed at two different exposure concentrations. Grey lines show 15 random simulation replicates, red dashed line indicates average across all simulations and solid black lines indicates the true simulated response.}
\label{fig:sim-slice-exp}
\centering
\end{figure}

\begin{figure}[!p]
\centering
\begin{subfigure}{.48\textwidth}
  \centering
  \includegraphics[width=\linewidth]{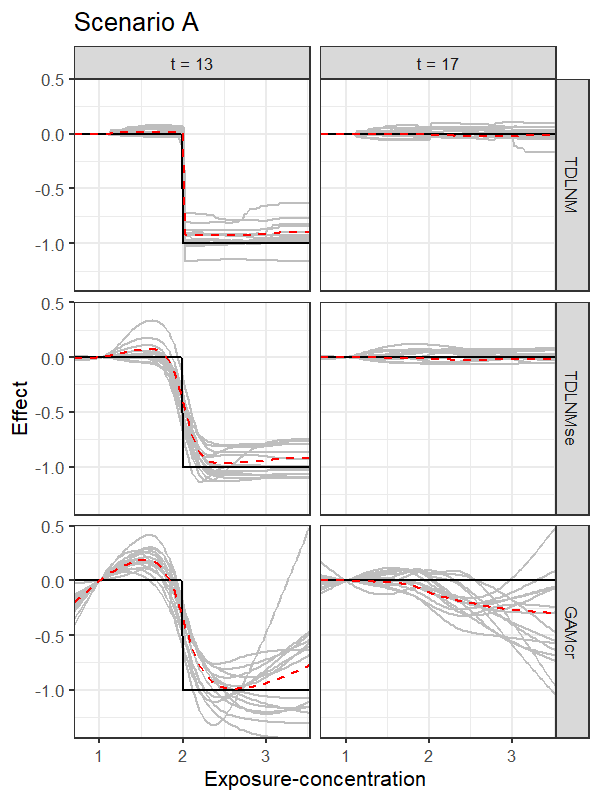}  
\end{subfigure}
\begin{subfigure}{.48\textwidth}
  \centering
  \includegraphics[width=\linewidth]{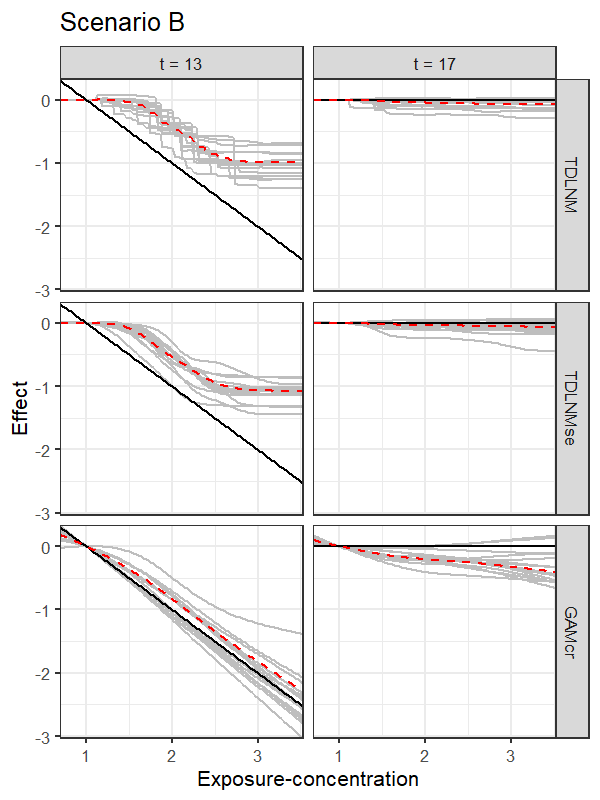}  
\end{subfigure}
\begin{subfigure}{.48\textwidth}
  \centering
  \includegraphics[width=\linewidth]{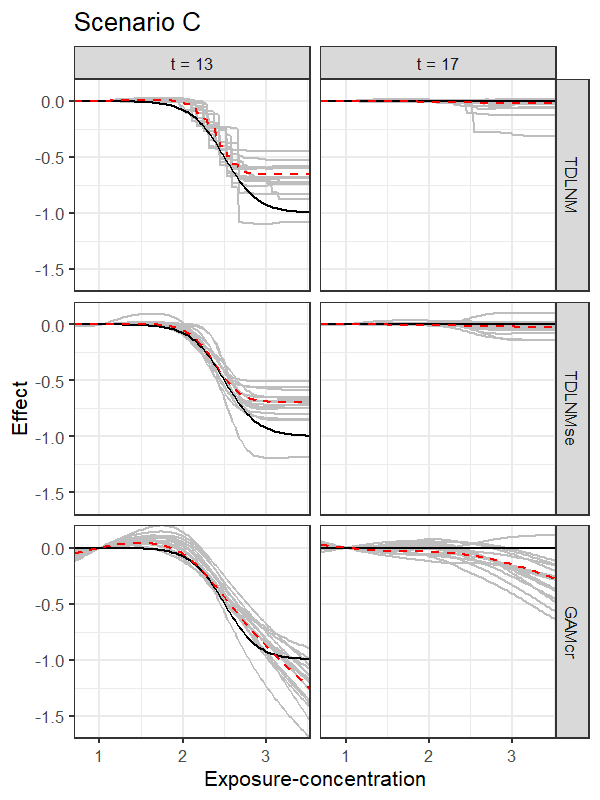}  
\end{subfigure}
\begin{subfigure}{.48\textwidth}
  \centering
  \includegraphics[width=\linewidth]{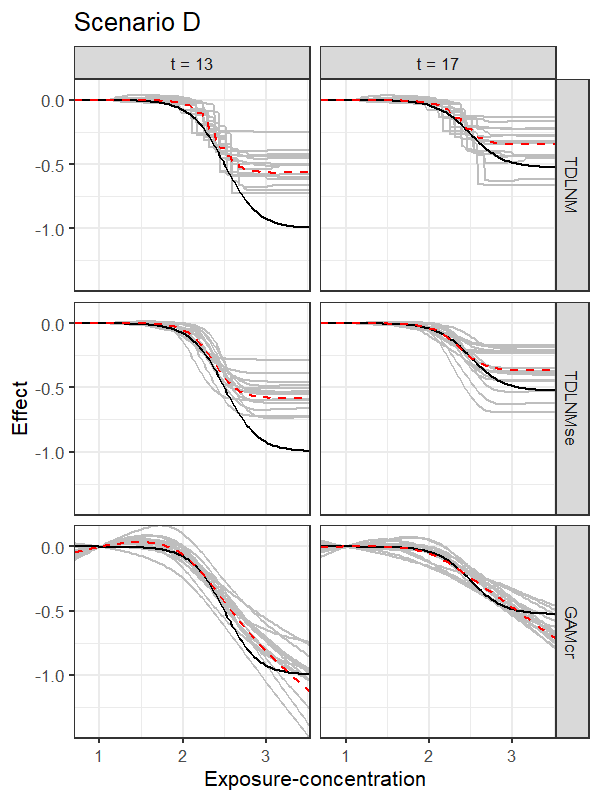}  
\end{subfigure}
\caption{Simulation results of TDLNM, TDLNMse, and GAMcr comparing the exposure-time-response (y-axis) cross-section across all exposure concentrations (x-axis) fixed at two different times. Grey lines show 15 random simulation replicates, red dashed line indicates average across all simulations and solid black lines indicates the true simulated response.}
\label{fig:sim-slice-time}
\centering
\end{figure}

TDLNM and TDLNMse have as good or better overall RMSE in scenarios A, C and D. In all scenarios, the tree-based methods have the lowest RMSE in areas of zero effect in the exposure-time-response surface. Figure \ref{fig:sim-slice-exp} highlights the ability of TDLNM and TDLNMse to find a sharp distinction between times with or without effects. The shrinkage prior on the tree-specific parameters reduces variance leading to lower RMSE in areas of no effect. In areas of non-zero effect, our models have lower RMSE than spline based models in scenario A and are comparable in scenarios C and D. In scenario B the RMSE in areas of non-zero effect is higher for TDLNM and TDLNMse, as the spline based models do a better job interpolating into the extreme exposure values where few data points reside. Figure \ref{fig:sim-slice-time} contrasts how tree-based models attenuate the effect at the boundaries of exposure values, while GAMcr continues the trend linearly.

The tree-based models have near nominal coverage, except in scenario B. All models show below nominal coverage in scenario B, however, TDLNM and TDLNMse perform best, each having 87\% surface coverage. In addition, our models have the smallest average confidence interval width, which is particularly notable at the boundaries in time or extreme exposure concentration where the `wiggliness' of spline-based models becomes more pronounced (Figures \ref{fig:sim-slice-exp} and \ref{fig:sim-slice-time}). The lack of `wiggliness' in the tree-based model estimates contributes to narrow confidence intervals as well as decreased RMSE, especially in areas of zero-effect. Furthermore, the variation between simulation replicates is much smaller for TDLNM and TDLNMse.

Scenario B, while seemingly natural for a DLM, poses several difficult situations. First, a proper estimate by TDLNM would require trees with many breaks spanning the exposure concentration during the correct critical window. Second, TDLNM attenuates the effect when data is sparse (e.g. high and low concentrations in this scenario). Third, at high concentration, there is a jump from zero to a large effect that smooth methods cannot accommodate; in particular, DLM extends the critical window well beyond the true period of effect as a result of the smoothness assumption.

Precision with TDLNM and TDLNMse is the highest across all simulation scenarios (Table \ref{tab:sim-results}). The high precision is a result of near zero FP, but with a tradeoff of lower TP in scenarios B and D. The cross-sectional plots in Figure \ref{fig:sim-slice-exp} shows the ability of TDLNM and TDLNMse to adapt to non-smooth exposure-time response surfaces. Supplemental Figure 3 indicates the probability detecting a non-zero effect in at least one exposure value in each week. These results shows that the spline-based methods have a much higher probability of misclassifying weeks just outside of the true critical windows. On the other hand, the tree-based models adapt to changing smoothness in the exposure-time-response surface and rarely detect non-zero effects outside of the true critical window. The key takeaway is that the critical windows detected by TDLNM and TDLNMse have a high probably of being correct.

\section{Data Analysis}
We use TDLNM and TDLNMse to estimate the relationship between a mother's exposure to PM$_{2.5}$ during the first 37 weeks of pregnancy and child BWGAZ. By using weekly exposures, we limit the temporal resolution at which critical windows can be identified with any method to correspond to weeks. For comparison, we also apply DLNM using penalized cubic regression splines (GAMcr) and DLM. We control for maternal baseline characteristics and season and long-term trends. The maternal characteristics are: pre-pregnancy age (quadratic fit), weight, smoking (if done before or during pregnancy), income, education, prenatal care (when first received), race and Hispanic designations, elevation, and county of residence. We do not control for fetal sex or gestational age as the outcome, BWGAZ, is already adjusted for these factors. In addition, we adjust for seasonal effects using indicators for year and month of conception.

For TDLNM and TDLNMse, we use the same hyperparameters as in our simulation, running the models for a burn-in period of $5,000$ iterations followed by $15,000$ iterations retaining every tenth draw from our MCMC sampler. We specify 30 equally spaced potential splits in the exposure dimension ranging from the 0.1 percentile to the 99.9 percentile of log-exposure values. Different numbers of potential splits were considered, but showed no differences in the result. In TDLNMse we set the smoothing parameter $\sigma_x$ equal to half the standard deviation of the log-exposures. Models GAMcr and DLM used the same settings as in simulation. The DLNM estimates for all models are centered at the median exposure value (approximately 7 $\mu$g/m$^3$). Critical windows are defined as any week containing a region in the exposure-time-response where the 95\% confidence interval does not contain zero.

\subsection{DLNM Results}
\begin{figure}[!p]
\centering
\begin{subfigure}{\textwidth}
  \centering
  \includegraphics[width=.95\linewidth]{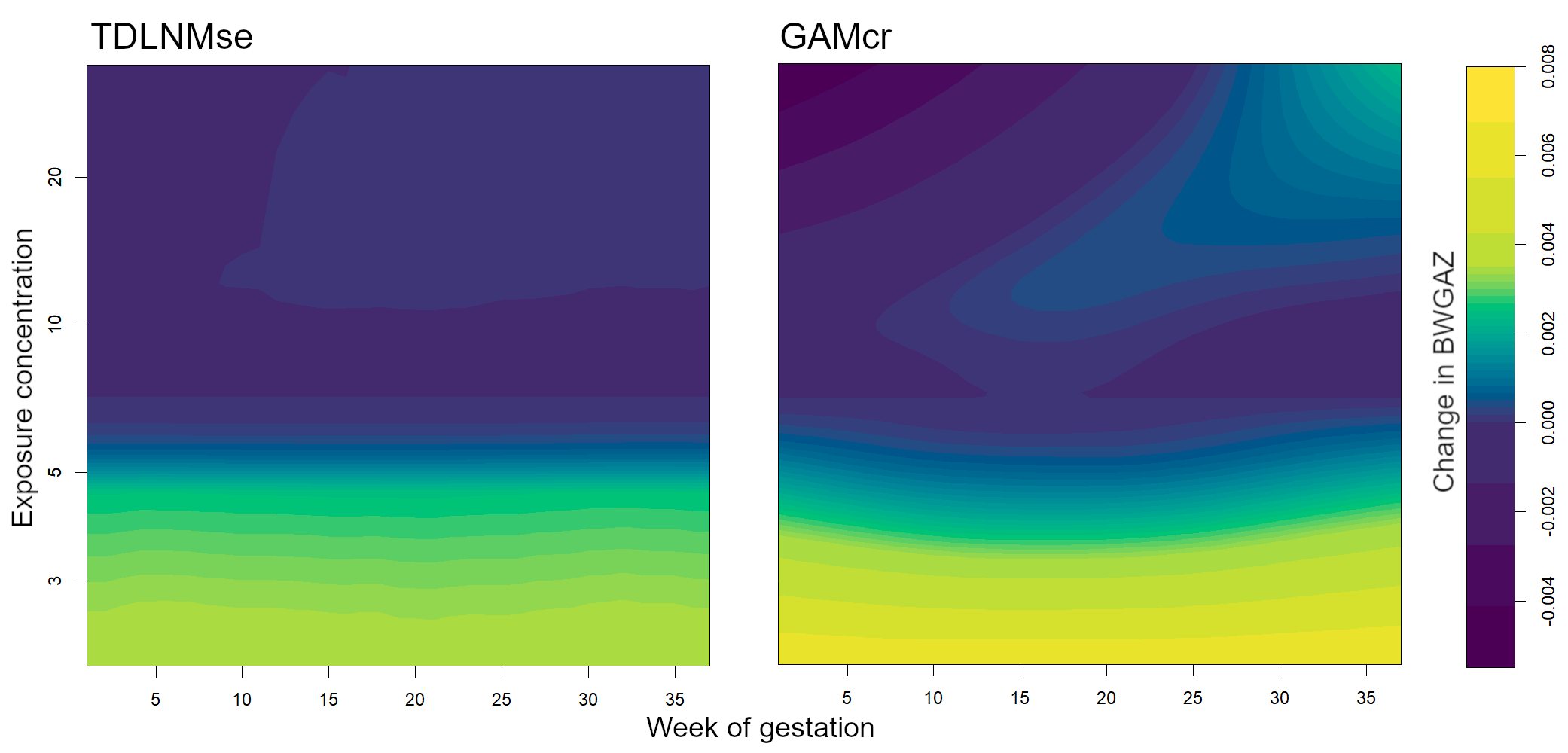}  
  \caption{Exposure-time-response surface}
  \label{fig:data-analysis-surface}
\end{subfigure}
\begin{subfigure}{\textwidth}
  \centering
  \includegraphics[width=.95\linewidth]{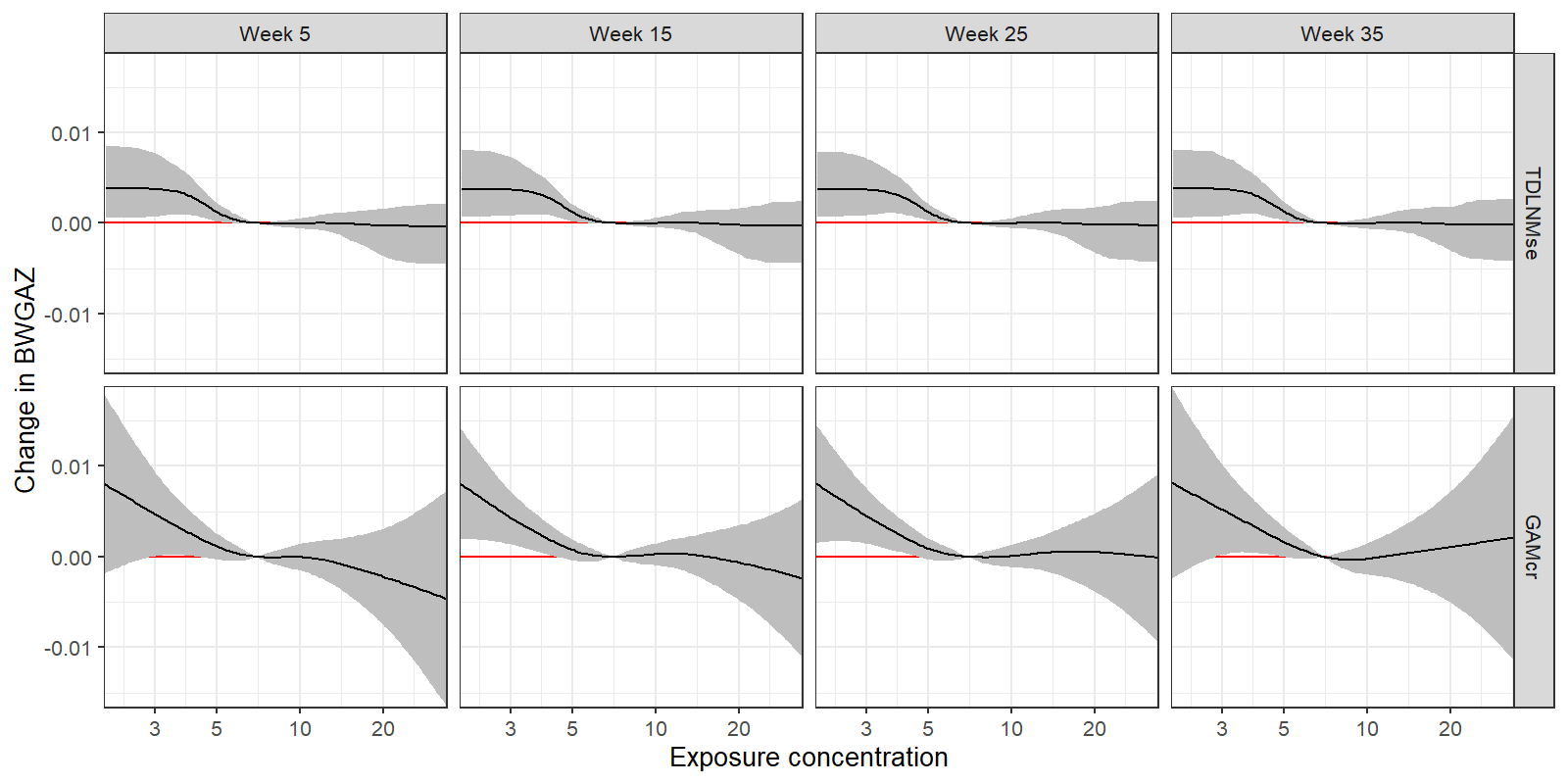} 
  \caption{Exposure-response function at weeks 5, 15, 25, and 35}
  \label{fig:data-analysis-slice}
\end{subfigure}
\caption{Panel (a) shows the estimated exposure-time-response surface for models TDLNMse and GAMcr. Panel (b) shows cross sections of the estimated exposure-time-response with columns showing the estimated effect at four times: $t=5,15,25,35$, while rows compare models TDLNMse and GAMcr. All plots indicate the exposure-time-response relative to the median exposure concentration value (7.0$\mu g/m^3$).}
\label{fig:data-analysis-mean}
\centering
\end{figure}

The posterior mean exposure-time-response estimates for TDLNMse is shown in Figure \ref{fig:data-analysis-surface}. PM$_{2.5}$ exposure below the median is associated with an increase in BWGAZ. Exposure concentration above the median value indicate a slight decrease in BWGAZ, but the 95\% credible intervals do not give reason to believe this is different from zero. This pattern is present across all gestational weeks. Cross-sections of the exposure-time-response surface at weeks 5, 15, 25, and 35 are shown in Figure \ref{fig:data-analysis-slice} and indicate a critical window spanning the entire pregnancy.

Based on TDLNMse, a change from median (7.0 $\mu g/m^3$) to the 25th percentile of PM$_{2.5}$ exposure (5.89 $\mu g/m^3$) across the pregnancy would result in a cumulative mean increase in BWGAZ of 0.0132 (95\% CI $[0.0003, 0.0354]$) or approximately 5.74g (95\% CI $[0.11, 15.41]$) when translated to actual birth weight (this is approximate because BWGAZ accounts for gestational age and fetal sex). The nonlinear association shows that a further decrease in  PM$_{2.5}$ exposure to the 10th percentile (5.02 $\mu g/m^3$) would result in a 0.055 (95\% CI $[0.016, 0.090]$) mean increase in BWGAZ, or an approximate increase of 24.1g (95\% CI $[7.155, 39.10]$). These results suggest that decreasing PM$_{2.5}$ below the current national ambient air quality standards would result in higher average birth weights in this population.

The mean exposure-time-response estimate for GAMcr, shown in Figure \ref{fig:data-analysis-surface}, closely resembles the estimates of TDLNMse. As in our simulations, we see a difference in the tail behavior. GAMcr continues the trend in the effect with large intervals. Despite the large point estimate with GAMcr at low exposure levels the larger confidence intervals include zero. In contrast, TDLNMse tapers off and estimates a smaller effect with substantially smaller intervals that do not contain zero. The smaller intervals found in TDLNMse near the boundaries are a result of these boundary regions being grouped in terminal nodes that also contain internal regions and therefore receive the same estimates.

Our findings of an association between increased PM$_{2.5}$ and decreased BWGAZ are consistent with previous literature. A meta-analysis by \cite{Sun2016TheMeta-analysis} found a 10 $\mu$g/$m^3$ increase in PM$_{2.5}$ across pregnancy to be associated with 15.9g decrease in birth weight (95\% CI $[-26.8,-5]$); increased exposures in the second and third trimesters were also determined to have a nonzero negative association with birth weight. \cite{Zhu2015a} reported similar results in a separate meta-analysis. \cite{Strickland2019} found that the magnitude of associations between PM$_{2.5}$ and birth weight increased for higher percentiles of the birth weight distribution across all trimesters. Finally, a study investigating individual chemical components of PM$_{2.5}$ found non-zero increased risk of low birth weight for maternal exposures during each trimester of pregnancy \citep{Ebisu2012}. 

\subsection{Comparing less flexible model alternatives}

For comparison, we fit TDLNM, a DLM and several linear models to compare results. Each of these models was consistent with the TDLNMse results. More details on these methods can be found in Supplemental Materials Section D.

\section{Discussion}

In this work we have proposed a tree-based method for a DLNM to estimate the association between a time-resolved series of pollution exposures and a continuous birth outcome. TDLNM eliminates the smoothness assumption in the exposure-time response surface. TDLNMse imposes smoothness only in the exposure-concentration dimension but not over time. TDLNM also has the potential to account for measurement error within the exposure-response function. By relaxing the smoothness assumption in the time dimension, our new methods more precisely identify critical windows of susceptibility.

TDLNM provides several extensions to tree-based regression models. First, we allow for a multivariate predictor with temporal correlation. Second, we provide a computationally efficient method for estimating a tree-based function while controlling for a fixed effect. Finally, we eliminate the need for cross-validation to select variance hyperpriors through the application of a horseshoe-like prior on tree-specific effects.

In simulation scenarios, we show that TDLNM and TDLNMse have a low false positive rate of critical window identification, while spline-based DLNMs have a tendency to over-generalization the time periods containing critical windows. Furthermore, our tree-based methods can approximate both smooth and non-smooth exposure-time-response functions. As the smoothness assumption in time changes, TDLNM and TDLNMse allow for information sharing at the same exposure levels across time, so that the piecewise constant steps are distributed across adjacent times allowing for near-smooth estimates. The shrinkage priors reduce the variance of estimates, reducing RMSE in areas of no effect and decreasing the rate of false positives. In the presence of a linear trend, DLNM models are overly flexible. While penalized spline DLNM can revert to an approximately linear model, TDLNM requires a large number of splits in the exposure-concentration dimension to accomplish the same results. As seen in simulation Scenario B, TDLNMse attenuated the linear trend in areas of few exposure observations. The simulations indicate that TDLNM and TDLNMse have high precision in identifying critical windows.

We applied TDLNM and TDLNMse to a Colorado birth cohort. We found a nonlinear effect of PM$_{2.5}$ on BWGAZ. Specifically, we found that below median levels of PM$_{2.5}$ throughout pregnancy were associated with higher BWGAZ. We found no change in BWGAZ due to above median PM$_{2.5}$ exposure.

\section*{Supplementary Materials}
The reader is referred to online Supplementary Materials for technical appendices and additional results concerning simulation and data analysis. The {\tt R} package {\tt dlmtree} used in the simulation and data analysis can be found at \url{https://github.com/danielmork/dlmtree}.

\section*{Acknowledgement}

This work was supported by National Institutes of Health grant ES028811.

These data were supplied by the Center for Health and Environmental Data Vital Statistics Program of the Colorado Department of Public Health and Environment, which specifically disclaims responsibility for any analyses, interpretations, or conclusions it has not provided.

This work utilized the RMACC Summit supercomputer, which is supported by the National Science Foundation (awards ACI-1532235 and ACI-1532236), the University of Colorado Boulder and Colorado State University. The RMACC Summit supercomputer is a joint effort of the University of Colorado Boulder and Colorado State University.

\bibliography{ref}

\clearpage

\begin{table}[!p]
\footnotesize
\setstretch{1.2}
\centering
\begin{tabular}{lrrrrrrrr}
  \toprule[2pt]
  &\multicolumn{3}{c}{DLNM RMSE}&
  \multicolumn{2}{c}{DLNM Coverage}&\multicolumn{3}{c}{Effect Identification}\\
  \cmidrule(lr){2-4} \cmidrule(lr){5-6} \cmidrule(lr){7-9}
  Model& Overall & No Effect & Effect & Overall & CI Width & TP & FP & Precision \\ 
  \hline
  \multicolumn{9}{l}{\textit{Scenario A: Piecewise in Exposure and Time}}\\
  \tt{TDLNM}    & 0.086 & 0.066 & 0.213 & 1.00 & 0.43 & 0.87 & 0.00 & 1.00 \\
  \tt{TDLNMse}  & 0.100 & 0.077 & 0.252 & 0.99 & 0.46 & 0.82 & 0.00 & 0.98 \\
  \tt{GAM}      & 0.294 & 0.258 & 0.584 & 0.95 & 1.08 & 0.47 & 0.03 & 0.90 \\
  \tt{DLM}      & 0.370 & 0.342 & 0.626 & 0.68 & 0.53 & 1.00 & 0.30 & 0.77 \\
  \tt{GLM-AIC}  & 1.531 & 1.536 & 1.462 & 0.84 & 3.35 & 0.49 & 0.15 & 0.55 \\
  \tt{GAMcr}    & 0.263 & 0.241 & 0.454 & 0.98 & 1.10 & 0.62 & 0.01 & 0.96 \\
  \tt{GAM-exp}  & 0.241 & 0.165 & 0.669 & 0.94 & 0.67 & 0.32 & 0.01 & 0.87 \\ 
  \multicolumn{9}{l}{\textit{Scenario B: Linear in Exposure }}\\
  \tt{TDLNM}    & 0.292 & 0.081 & 0.768 & 0.87 & 0.37 & 0.56 & 0.01 & 0.99 \\ 
  \tt{TDLNMse}  & 0.270 & 0.073 & 0.712 & 0.87 & 0.34 & 0.64 & 0.01 & 0.99 \\
  \tt{GAM}      & 0.312 & 0.257 & 0.547 & 0.73 & 0.48 & 0.90 & 0.18 & 0.84 \\
  \tt{DLM}      & 0.299 & 0.257 & 0.489 & 0.64 & 0.36 & 1.00 & 0.26 & 0.79 \\
  \tt{GLM-AIC}  & 0.267 & 0.253 & 0.346 & 0.79 & 0.46 & 0.99 & 0.18 & 0.85 \\
  \tt{GAMcr}    & 0.248 & 0.206 & 0.426 & 0.84 & 0.54 & 0.87 & 0.09 & 0.90 \\
  \tt{GAM-exp}  & 0.283 & 0.226 & 0.518 & 0.76 & 0.37 & 0.94 & 0.15 & 0.86 \\ 
  \multicolumn{9}{l}{\textit{Scenario C: Smooth in Exposure}}\\
  \tt{TDLNM}    & 0.077 & 0.033 & 0.223 & 0.94 & 0.18 & 0.58 & 0.01 & 0.99 \\ 
  \tt{TDLNMse}  & 0.070 & 0.032 & 0.201 & 0.97 & 0.17 & 0.67 & 0.01 & 0.99 \\ 
  \tt{GAM}      & 0.142 & 0.126 & 0.241 & 0.91 & 0.36 & 0.60 & 0.06 & 0.91 \\ 
  \tt{DLM}      & 0.138 & 0.120 & 0.245 & 0.64 & 0.18 & 1.00 & 0.31 & 0.77 \\ 
  \tt{GLM-AIC}  & 0.186 & 0.167 & 0.309 & 0.82 & 0.40 & 0.53 & 0.14 & 0.80 \\ 
  \tt{GAMcr}    & 0.113 & 0.104 & 0.176 & 0.95 & 0.37 & 0.64 & 0.03 & 0.96 \\ 
  \tt{GAM-exp}  & 0.126 & 0.103 & 0.255 & 0.92 & 0.28 & 0.62 & 0.05 & 0.93 \\ 
  \multicolumn{9}{l}{\textit{Scenario D: Smooth in Exposure and Time}}\\
  \tt{TDLNM}    & 0.105 & 0.041 & 0.203 & 0.80 & 0.26 & 0.40 & 0.00 & 0.99 \\ 
  \tt{TDLNMse}  & 0.098 & 0.038 & 0.190 & 0.95 & 0.24 & 0.45 & 0.01 & 0.99 \\ 
  \tt{GAM}      & 0.120 & 0.100 & 0.171 & 0.97 & 0.44 & 0.54 & 0.01 & 0.98 \\ 
  \tt{DLM}      & 0.122 & 0.090 & 0.193 & 0.69 & 0.23 & 0.94 & 0.23 & 0.80 \\ 
  \tt{GLM-AIC}  & 0.284 & 0.277 & 0.306 & 0.81 & 0.52 & 0.45 & 0.14 & 0.77 \\ 
  \tt{GAMcr}    & 0.110 & 0.092 & 0.156 & 0.97 & 0.41 & 0.57 & 0.01 & 0.98 \\ 
  \tt{GAM-exp}  & 0.099 & 0.068 & 0.164 & 0.97 & 0.35 & 0.47 & 0.00 & 0.99 \\ 
  \bottomrule[2pt]
\end{tabular}
\caption{Simulation results, showing RMSE for estimation of the exposure-time-surface with no-effect and effect separated. Coverage and CI width is based on 95\% confidence intervals. Effect identification considers the likelihood of identifying a non-zero effect (TP) or incorrectly designating a non-zero effect (FP) over the DLNM surface. Precision is calculated within each simulation as TP/(TP+FP). Standard errors are available in Supplemental Materials Table 2.}
\label{tab:sim-results}
\end{table}

\end{document}